\documentclass[journal]{IEEEtran}
\usepackage{graphicx} 
\usepackage{amsmath} 
\usepackage{subcaption}
\usepackage{cite}
\begin{document}

\title{Deep Semantic Segmentation for Multi-Source Localization Using Angle of Arrival Measurements}

\author{Mustafa~Atahan~Nuhoglu~\IEEEmembership{}
	and~Hakan~Ali~Cirpan,~\IEEEmembership{Member,~IEEE}
	\thanks{M.A. Nuhoglu is with Radar, Electronic Warfare and Intelligence Systems Division, ASELSAN A.S., Ankara, 06830, Turkey. e-mail: manuhoglu@aselsan.com.tr}
	\thanks{M.A. Nuhoglu and H.A. Cirpan are with Istanbul Technical University, Istanbul, 34467, Turkey. e-mail: cirpanh@itu.edu.tr}
}
\maketitle

\begin{abstract}
This paper presents a solution for multi-source localization using only angle of arrival measurements. The receiver platform is in motion, while the sources are assumed to be stationary. Although numerous methods exist for single-source localization—many relying on pseudo-linear formulations or non-convex optimization techniques—there remains a significant gap in research addressing multi-source localization in dynamic environments. To bridge this gap, we propose a deep learning-based framework that leverages semantic segmentation models for multi-source localization. Specifically, we employ U-Net and U-Net++ as backbone models, processing input images that encode the platform's positions along with the corresponding direction-finding lines at each position. By analyzing the intersections of these lines, the models effectively identify and localize multiple sources. Through simulations, we evaluate both single- and multi-source localization scenarios. Our results demonstrate that while the proposed approach performs comparably to traditional methods in single-source localization, it achieves accurate source localization even in challenging conditions with high noise levels and an increased number of sources.
\end{abstract}

\section{Introduction}

Source localization is a fundamental problem in aerospace and electronic engineering, with significant implications for electronic warfare (EW) systems, radar systems, sonar systems, and various wireless communication applications \cite{o2019emitter,hassan2021state,yen20223}. Traditional source localization techniques often rely on various types of measurements, with angle of arrival (AoA) data being one of the most utilized for systems where range measurements are either impractical or unavailable. Source localization based on AoA requires triangulation, which is usually employed by multiple stationary platforms positioned at different locations or by a single moving platform \cite{aubry2024sensor,widdison2024review,10778573,10018916}.  This provides the angular aperture needed for an unambiguous resolution of the source location. Single-source localization has been extensively studied, and several approaches have been proposed that involve pseudo-linear transforms and instrumental variables \cite{nardone1984fundamental,douganccay2004passive}.  While these strategies have shown promising results for single-source cases, they fall short when applied to multi-source localization. In scenarios where multiple sources emit signals simultaneously, the problem becomes significantly more complex as several key challenges emerge. 
First, the measurements collected by the system often contain contributions from multiple sources, requiring an effective approach to separate and accurately associate them with the correct sources. Second, the presence of ambiguous or overlapping  AoA measurements further complicates this process, as it becomes difficult to distinguish individual sources when their signals intersect or are closely spaced \cite{flood2024multi,lin2024hybrid}. Finally, if the receiver platform is in motion, the dynamic nature of its trajectory introduces additional complexities to both the measurement model and the optimization process, making it more challenging to achieve precise localization \cite{mansourian2024sparsity}.

Deep semantic segmentation models have lately garnered quite a bit of attention in the EW domain. They interpret complex patterns in spatial data, which enables systems to distinguish between different objects or regions within a given environment. Therefore, deep semantic segmentation enhances situational awareness and decision-making capabilities. One key advantage of these models is their effectiveness in problems without prior information. Note that EW systems mostly operate passively without any information about the incoming signal, and their usefulness in the EW domain has been proven by various works. These studies show prominent results compared to traditional methods in the deinterleaving problem, activity detection, and low probability of intercept radar detection \cite{nuhoglu2022image,9999169,10445810,10150837,10393280,9623058,10815794}. 

To the best of our knowledge, there is no existing solution that efficiently addresses the multi-source localization problem using AoA measurements in dynamic settings with no available prior information regarding sources. Here, we propose a deep learning framework using semantic segmentation for multi-source localization. The advantages of using deep learning models for this task are numerous. First, deep learning models, particularly convolutional neural networks (CNNs) used for semantic segmentation, are capable of learning complex mappings from input data to output labels without the need for manual feature engineering \cite{long2015fully,li2021survey}. This capability is essential for handling the non-linear nature of AoA measurements and the dynamic movement of the receiver platform. Second, deep learning-based segmentation models can efficiently process large volumes of data, making them suitable for real-time localization in practical applications \cite{peng2020deep,yu2018bisenet,siam2018comparative}. Additionally, they can interpret complex patterns in spatial data, making them capable of identifying intersections or convergence points \cite{mo2022review,yuan2021review,truong2021deep,knap2023boosted}. In our work, we employ two state-of-the-art deep learning architectures, UNet and UNet++, as the backbone of our localization model \cite{ronneberger2015u,zhou2018unet++}. Their key innovation lies in the use of a symmetric encoder-decoder structure, where the encoder captures the context, and the decoder enables precise localization. UNet’s architecture features skip connections that link corresponding layers in the encoder and decoder, allowing the network to retain spatial information during the segmentation process. This ensures precise segmentation of fine-grained features, crucial for pixel-level classification applications. The proposed approach processes input images containing the platform’s positions and the corresponding direction-finding (DF) lines at each position. These DF lines are constructed from AoA measurements obtained at different receiver locations. The intersections of these DF lines in the image correspond to potential source locations. By training the semantic segmentation models to identify these intersections, we can accurately localize the sources. Unlike existing optimization-based approaches, which are designed to handle only single-source localization, our method can handle scenarios with multiple sources and complex measurement patterns, providing a robust and scalable solution.

Our contributions to the source localization literature are summarized as follows:
\begin{enumerate}
    \item The development of a novel preprocessing method that transforms AoA measurements into two-dimensional images. 
    \item The development of a novel semantic segmentation-based framework for multi-source localization, trained on the preprocessed images.
    \item An evaluation of the performance of these models in terms of their ability to accurately localize a single source with dynamic receiver platforms.
    \item An evaluation of the performance of these models in terms of their ability to accurately localize multiple sources and handle the ambiguities associated with dynamic receiver platforms.
\end{enumerate}

The remainder of the paper is structured as follows: Section II reviews existing literature on source localization using AoA measurements, with a particular focus on single-source localization techniques and the challenges associated with multi-source scenarios. Section III details the methodology of our proposed semantic segmentation approach, including data preparation, model architectures, and training strategy. Sections IV and V present complexity analysis and experimental results, including model performance metrics and a comparative analysis against existing methods. Finally, Section VI concludes the paper with a discussion of the findings and potential avenues for future work.
\section{Signal Model and Related Work}
Source localization using AoA measurements has been a critical area of research in EW systems. Accurate localization of sources is essential for tasks like tracking, surveillance, and signal processing in dynamic environments. This section reviews the literature on AoA-based localization methods, focusing on single-source and multi-source scenarios, and discussing the limitations of existing methods in handling multi-source localization problems. In our work, we assume that the source is stationary and the platform carrying the EW receiver is in motion. 

Platform position error is a recognized factor that can introduce bias and degrade localization accuracy by shifting the perceived position of the platform away from its actual location, consequently impacting AoA measurements \cite{ren2024efficient,ho2007source,jia2024sensor}. However, to simplify our analysis, we consider a scenario without platform position errors, allowing us to focus solely on evaluating the performance of the proposed approach. 
\subsection{Single Source Localization}
Single-source localization involves determining the position \((x_p, y_p)\) of a single source based on measurements taken from a receiver platform that moves along a known trajectory. Here, the receiver positions are denoted by \((x_n, y_n)\), where \(n\) is the sample index. The AoA measurement is given by
\begin{equation}
\theta_n = \tan^{-1} \left( \frac{y_p - y_n}{x_p - x_n} \right)~.
\label{mlTheta}
\end{equation}

This relationship defines the direction vector from the receiver to the source, and it is non-linear due to the tangent function. Assume that the AoA measurements are corrupted by additive white Gaussian noise with zero mean and $\sigma^2$ variance. Then, the noisy measurements become
\begin{align}
    \tilde{\theta}_n = \theta_n  + \epsilon_n~,
\end{align}
where $\epsilon_n$ is the noise. In the single-source localization case, we categorize the existing methods into three: maximum likelihood, pseudo-linear least squares, and weighted instrumental variable estimators. They aim to estimate the source location vector $\mathbf{u} = [x_p,y_p]^T$ from AoA measurements that are collected over a time.

\subsubsection{Maximum Likelihood (ML)}
For single-source localization, the problem can be posed as an optimization task where the goal is to find the source position that minimizes the error across all AoA measurements. One common approach is to solve a least-squares optimization problem:
\begin{equation}
\hat{\mathbf{u}} = \arg\min_{x_p,y_p} \sum_{n=1}^{N} \left( \tilde{\theta}_n - \tan^{-1} \left( \frac{y_p - y_n}{x_p - x_n} \right) \right)^2~,
\end{equation}
where \(N\) is the total number of measurements. This problem is non-convex and non-linear in terms of the source location due to the tangent function and fraction operation. Therefore, it is not straightforward. Solutions to this problem include gradient-based methods such as Newton-Raphson, Gauss-Newton, or gradient descent \cite{zou2020novel,ahmed2020localization}. These methods can converge fast but they require proper initial points, which is usually not available in real-life scenarios. Additionally, since they are derivative-driven, they can get stuck in local minima. Another approach is the use of non-convex optimizers, which can handle any type of cost function. Grid search, genetic algorithms, and particle swarm optimization methods are well-known non-convex optimizers \cite{kennedy1995particle,gen1999genetic}. These methods do not require initial points and, with the right configurations, can avoid getting stuck in local minima. However, they are computationally expensive methods due to their iterative nature \cite{nuhoglu2020iterative,nuhoglu2024source}. 

\subsubsection{Pseudo-Linear Least Squares (PLS)}
PLS techniques have been widely used for solving single-source localization problems due to their simplicity and effectiveness. These methods are particularly useful when dealing with the non-linear nature of AoA measurements. The basic idea behind pseudo-linear least squares is to linearize the non-linear equations governing the AoA measurements. Taking the tangent of both sides in \eqref{mlTheta} results in
\begin{align}
    \tan(\theta_ n) &= \frac{y_p - y_n}{x_p - x_n}~, \\
    x_p\tan(\theta_ n)-y_p&= x_n\tan(\theta_ n)- y_n~, 
\end{align}
which forms a linear model as follows:
\begin{align}
\mathbf{b} = \mathbf{A}\mathbf{u}~.
\label{ls}
\end{align}
Here, 
\begin{align}
\mathbf{b} &= [b_1,b_2,...,b_N]^T~,\\
\mathbf{A} &= [\mathbf{A}_{1},\mathbf{A}_{2},...,\mathbf{A}_{N}]^T~,
\end{align}
where $b_n = x_n\tan(\theta_ n)- y_n$, $\mathbf{A_{n}=[\tan(\theta_ n), -1]^T}$. Then, the optimal solution to the measurement model in \eqref{ls} is obtained by the least-squares technique:
\begin{align}
\hat{\mathbf{u}} = (\mathbf{A}^T\mathbf{A})^{-1}\mathbf{A}^T\mathbf{b}~.
\label{PLS}
\end{align}
One significant drawback of these methods is their inherent high bias. This bias arises from the linearization of the non-linear model and the existing correlation between $\mathbf{A}$ and measurement noise, which can lead to suboptimal estimates of source positions. To address this issue and improve the accuracy of localization, the weighted instrumental variable estimator was proposed as an effective solution \cite{douganccay2004passive}.
\subsubsection{Weighted Instrumental Variable Estimator (WIVE)}

The WIVE approach enhances the pseudo-linear method by incorporating instrumental variables (IV) that are less sensitive to measurement noise and are chosen to provide more accurate representations of the true source positions. These instrumental variables help reduce the bias in the estimation process by ensuring that the residuals in the objective function are minimized in a weighted manner, taking into account the variability of the data. Using the estimated source location $\hat{\mathbf{u}} = [\hat{x}_p, \hat{y}_p]^T$ from \eqref{PLS} by the PLS method, AoA values are estimated as 
\begin{align}
\hat{\theta}_n &=  \tan^{-1} (\frac{\hat{y}_p - y_n}{\hat{x}_p - x_n})~.
\end{align}
By using the estimated AoA values, the IV matrix is formed as 
\begin{align}
\mathbf{G} &= [\mathbf{G}_{1},\mathbf{G}_{2},...,\mathbf{G}_{N}]^T~,
\end{align}
where $\mathbf{G_{n}=[\tan(\hat{\theta}_n), -1]^T}$. Then, the WIVE solution is obtained as
\begin{align}
\hat{\mathbf{u}} = (\mathbf{G}^T\mathbf{W}\mathbf{A})^{-1}\mathbf{G}^T\mathbf{W}\mathbf{b}~,
\label{WIVE}
\end{align}
where $\mathbf{W}=\operatorname{diag}\left(\hat{R}_1^2 \sigma_1^2, \hat{R}_2^2 \sigma_2^2, \ldots, \hat{R}_N^2 \sigma_N^2\right)$ is the weighting matrix, $\hat{R}_t=\sqrt{\left(\hat{x}_p-x_n\right)^2+\left(\hat{y}_p-y_n\right)^2}$ is the estimated range to the source which is calculated by using $\hat{\mathbf{u}} = [\hat{x}_p, \hat{y}_p]^T$ from \eqref{PLS} by the PLS method, and $\sigma$ is the standard deviation of the AoA measurement noise. The $\sigma$ values can be determined through an offline procedure. This involves collecting multiple AoA measurements while varying parameters such as frequencies, signal strengths, and relative directions. From these measurements,  AoA estimation errors are computed by comparing them to the true AoA values known from the simulation setup. This process creates a mapping between these parameters and $\sigma$ values. Using this offline mapping, which is stored in a look-up table, $\sigma$ values are retrieved online during the source localization process. Note that WIVE prioritizes measurements taken when the platform is closer to the source and when the noise level is lower, ensuring that more informative measurements have a greater influence on the solution. By appropriately selecting weights and instrumental variables, WIVE effectively reduces the bias introduced by the pseudo-linear approximation, resulting in a more robust and accurate solution for source localization. 

\subsection{Multi Source Localization}
The multi-source localization problem involves estimating the positions of \(S\) sources, \(\mathbf{u}^s = (x_p^s, y_p^s)\) for \(s = 1, 2, \ldots, S\), based on AoA measurements collected from the moving receiver at \(N\) positions. The measurements are expressed as:
\begin{equation}
\theta_n^s = \tan^{-1} \left( \frac{y_p^s - y_n}{x_p^s - x_n} \right),
\end{equation}
where \(\theta_n^s\) represents the AoA for the \(s\)-th source at the \(n\)-th measurement. There are two main issues that make the multi-source localization problem challenging. Firstly, there is no prior information regarding how many sources exist in the field. Therefore, $S$ is unknown. Secondly, it is also unknown which AoA measurement corresponds to which source. Hence, solving the problem via ML approaches or constructing a cost function that is based on a linear model is not possible.  Therefore, the single-source localization methods do not work in this case. To resolve the multi-source localization problem, we first develop a preprocessing technqiue to generate two-dimensional images that reveal the source locations. Then, we train state-of-the art deep learning models to achieve semantic segmentation, which eventually yields the source locations.  
\begin{figure}
   \centering
    \includegraphics[width=1\linewidth]{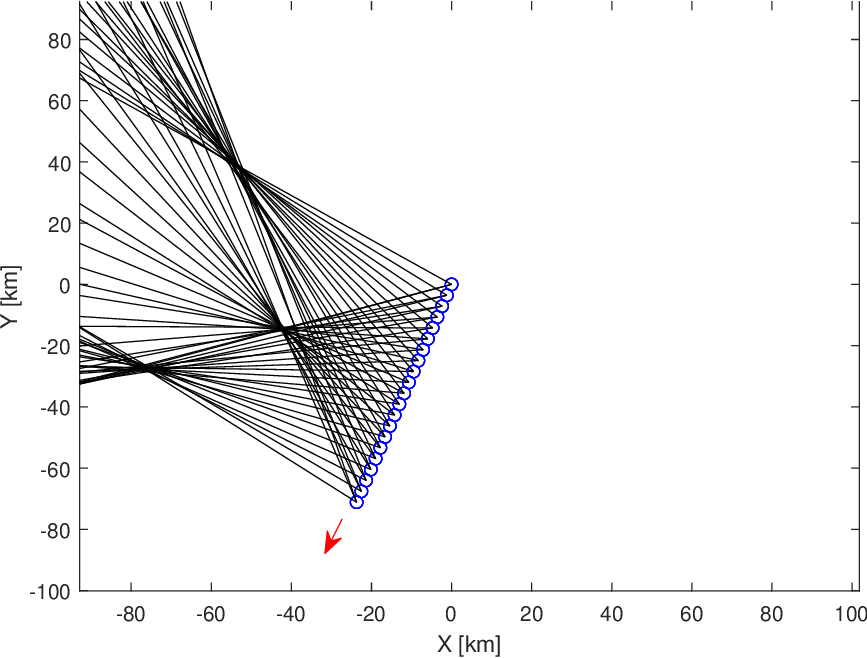}
    \caption{Multi-source localization scenario. Blue circles represent the platform positions and the red arrowhead indicates the direction of the movement. Black lines, which are the DF lines originating from the platform positions, intersect at source locations. }
   \label{fig:multiSource}
\end{figure}
 \begin{figure}[t!]
	\centering
	\includegraphics[width=1\linewidth]{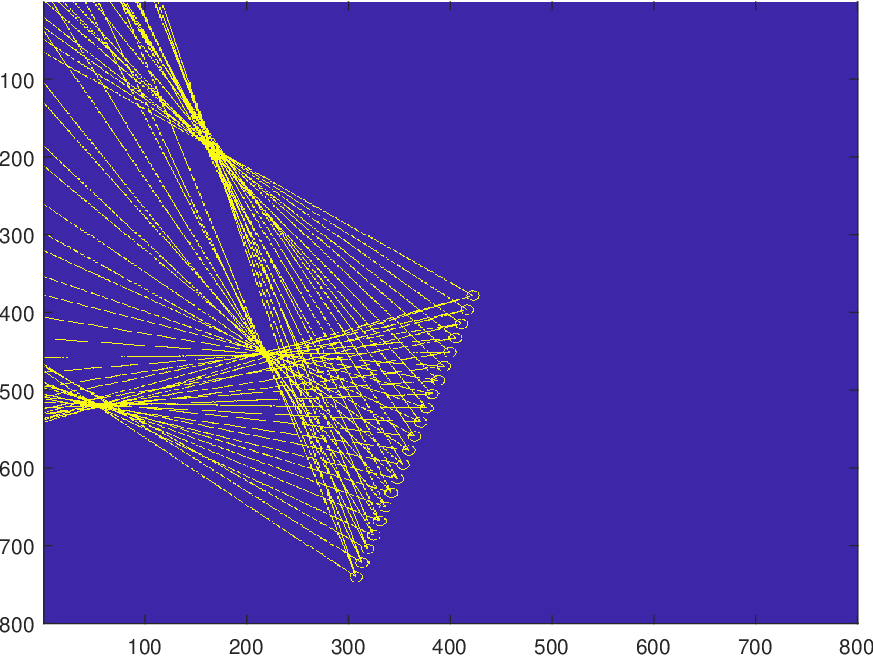}
	\caption{Generated input image. The platform, five source positions and DF lines are shown, when $\sigma$ $= 1$ degree. DF lines  intersect in the regions covering the source positions. }
	\label{fig:inputImage}
\end{figure}

\begin{figure}[t!]
	\centering
	\includegraphics[width=1\linewidth]{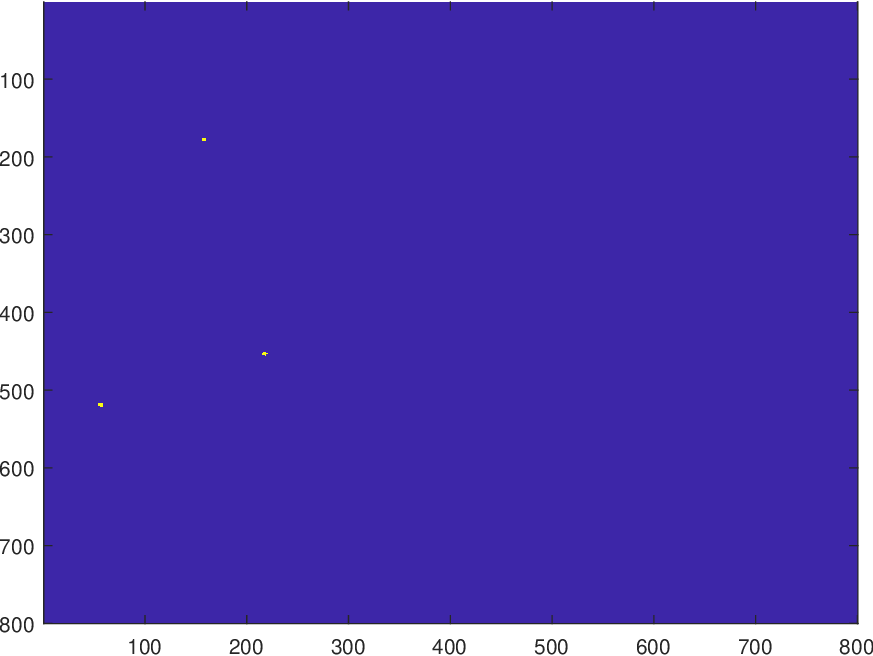}
	\caption{Generated label image. The yellow dots indicate true positions of the sources.}
	\label{fig:labelImage}
\end{figure}


\section{Proposed Approach}
First, we preprocess the noisy AoA measurements collected over time, transforming them into input images for the deep segmentation models. These images are then fed into the models, which output segmentation maps where distinct regions indicate possible source locations. Next, we apply connected component analysis (CCA) to identify and separate these regions. CCA is a technique that groups pixels based on connectivity, ensuring that pixels belonging to the same detected region are clustered together while different regions are distinguished \cite{dillencourt1992general,he2017connected}. This step is essential in differentiating multiple sources, as it prevents overlapping or adjacent detections from being treated as a single source. Following this, we perform centroid estimation to determine the central point of each detected region. The centroid is computed as the average of the pixel coordinates within a connected component, effectively providing a representative location for each source \cite{madsen1989estimating,zhao2025real}. Since the deep learning model’s output is in image space, these centroids correspond to pixel coordinates in the segmentation map. Finally, the estimated centroids are mapped from pixel space to real-world Cartesian coordinates, yielding the final source locations. 
\subsection{Proposed Preprocessing}

The multi-source localization problem is inherently mathematically challenging due to its non-convex nature and the complex interactions between the measurements and the unknown source locations. However, visualizing the problem makes it more tractable and easier to interpret. To facilitate this, we propose a novel preprocessing technique that transforms the multi-source localization task into a two-dimensional image format suitable for input into semantic segmentation models.

To generate the input images for the deep models, we first plot the multi-source localization scenario, as shown in Fig. \ref{fig:multiSource}. In this figure, platform positions are represented by blue circles, while the movement direction is indicated by a red arrowhead. Each DF line, corresponding to an AoA measurement, is depicted as a black line originating from the respective platform position. The scenario includes three sources, and the DF lines intersect within the regions containing these sources. Next, the scenario image is converted into a single-channel grayscale image and normalized. The resulting input image is visualized in Fig. \ref{fig:inputImage}. Additionally, we present the corresponding binary label image in Fig. \ref{fig:labelImage}, where true source locations are marked as dots. In this label image, pixels corresponding to source locations are set to one, while all other pixels are set to zero. The deep models are trained to map the input image to the label image by detecting the most probable intersection points of the DF lines. To further clarify the image generation and binarization process for input into the semantic segmentation model, we consider the following aspects:

\begin{enumerate}
    \item \textbf{Image Size and Resolution:}  
    The spatial region of interest covers a range of $[-100 \, \text{km}, +100 \, \text{km}]$ along both the X and Y axes, resulting in a total area of $200 \, \text{km} \times 200 \, \text{km}$. The range resolution is set to $250 \, \text{m}$, leading to a grid size of $800 \times 800$ pixels for each image.  The input images are normalized to ensure consistency in brightness and contrast, aiding the model's learning process.
    
    \item \textbf{Platform Representation:}  
    Each measurement platform's position is visualized as a circle in the image, with a fixed size that ensures visibility while not obscuring other elements. The circle is centered at the platform's location, defined by its coordinates mapped to the pixel grid.  
    
    \item \textbf{DF Line Drawing:}  
    From each platform position, DF lines are drawn based on the measured AoA values. These lines are extended outward to cover the entire image, ensuring they accurately represent the potential source directions.  
    
    \item \textbf{Binary Label Image Generation:}  
    The binary label image is also $800 \times 800$ pixels, matching the resolution of the input image. The source locations are represented as pixels with a value of one at the corresponding positions, while all other pixels are set to zero, resulting in a sparse binary representation.  
\end{enumerate}

Note that the sparsity of the binary label image arises because pixels indicating source locations are relatively few compared to the total number of pixels. Therefore, the number of foreground pixels is quite low compared to the number of background pixels. This characteristic is well-suited for semantic segmentation models, as it allows them to focus on identifying localized regions of interest at the intersections of DF lines. This structured image representation enhances the interpretability of the multi-source localization problem and facilitates accurate learning by the semantic segmentation model.

\subsection{Proposed Models}
We chose UNet and UNet++ as the backbone architectures for our multi-source localization framework due to their strong capability in learning spatial structures from limited labeled data. UNet's encoder-decoder structure, combined with skip connections, ensures high spatial resolution in segmentation tasks, which is crucial for accurately identifying source locations from intersecting DF lines. UNet++, as an extension of UNet, further enhances feature fusion through nested dense skip connections, improving robustness in scenarios where multiple sources are closely spaced or ambiguities arise due to overlapping DF lines.

Alternative segmentation architectures, such as transformer-based models, were also considered  \cite{chen2017rethinking,chen2018encoder, strudel2021segmenter,xie2021segformer}. For example, DeepLabV3+ employs atrous spatial pyramid pooling for multi-scale feature extraction, which is beneficial for segmenting objects at different scales. However, our problem primarily requires precise localization of structured intersection points rather than general scene parsing, making UNet-style architectures more suitable. Transformer-based segmentation models, while powerful for capturing long-range dependencies, tend to have higher computational complexity and inference time, which can be a limitation for real-time or resource-constrained applications. Given these considerations, UNet and UNet++ provide an optimal balance between segmentation accuracy, computational efficiency, and suitability for AoA-based localization.
\subsubsection{UNet}
UNet is a deep learning architecture primarily designed for semantic image segmentation tasks \cite{ronneberger2015u}. Initially introduced for biomedical image segmentation, it has since been widely adopted in various domains. The UNet architecture is characterized by its distinctive U-shaped structure, which consists of a contracting path (encoder) and an expanding path (decoder). The encoder captures context through down-sampling operations, while the decoder enables precise localization by up-sampling and integrating high-resolution features from the encoder via skip connections.

UNet is particularly effective for multi-source localization because it can learn the geometric structures present in AoA-derived images. In these images, the DF lines originating from different platform positions intersect at true source locations, forming structured patterns. The encoder extracts hierarchical spatial features, learning how individual DF lines contribute to potential source locations, while the decoder refines these representations to accurately classify source regions. The skip connections help retain fine-grained spatial details, ensuring that small but critical intersection regions are preserved in the final segmentation output.
\subsubsection{UNet++}
UNet++ is an advanced variant of the original UNet architecture, designed to improve the ability of deep learning models to capture intricate features and spatial relationships in image segmentation tasks \cite{zhou2018unet++}. The primary innovation of UNet++ lies in its nested and skip pathways, which enhance the model’s ability to propagate information across different layers, improving the network’s robustness and accuracy.

UNet++ further refines feature extraction for localization tasks by incorporating dense skip connections, which improve the fusion of low-level and high-level features. This is particularly useful for distinguishing closely spaced sources, as the additional convolutional layers in skip pathways help mitigate ambiguity in DF line intersections. The network learns to identify intersection points more precisely and reduces false alarms by refining feature propagation across scales. This makes UNet++ a strong candidate for handling complex multi-source localization scenarios where sources are closely spaced or DF lines exhibit significant overlap.

\section{Complexity Analysis}
To evaluate the computational feasibility of the proposed approach, we analyze the complexity of each method in terms of its fundamental operations. ML, PLS, and WIVE involve matrix operations, iterative solvers, and cost function evaluations, leading to varying computational demands. On the other hand, deep learning-based approaches like UNet and UNet++ utilize CNNs, which can be easily parallelized, for feature extraction and pattern recognition. This section provides a detailed breakdown of these complexities, comparing their efficiency and scalability in source localization tasks.
\subsection{ML}
We prefer the particle swarm optimization (PSO) algorithm to solve the ML cost function. The computational complexity of the PSO algorithm primarily depends on the number of particles \( M \) and the dimensionality of the search space \( K \). In each iteration, PSO evaluates the objective function for all \( M \) particles, updates their velocities and positions, and determines the global and local best solutions. In each iteration and particle, \eqref{mlTheta} is calculated, leading to a per-iteration complexity of \( O(NMK) \). Over \( T \) iterations, the total complexity of PSO is \( O(TNMK) \). The convergence rate and total iterations required depend on the problem characteristics, tuning parameters, and stopping criteria, but in general, PSO is computationally efficient compared to exhaustive search methods.

\subsection{PLS}
The computational complexity of the PLS method primarily depends on the size of the input matrix, which is determined by the number of measurements \( N \) and the number of variables \( K \). Given an \( N \times K \) measurement matrix \( A \), the standard approach involves solving the normal equations \( A^T A \mathbf{u} = A^T \mathbf{b} \), where \( \mathbf{u} \) is the vector of unknown parameters, and \( \mathbf{b} \) is the vector of observed values. The matrix multiplication \( A^T A \) has a computational complexity of \( O(NK^2) \), and solving the resulting system of equations typically requires an inversion or factorization, which takes \( O(K^3) \) operations. Therefore, the overall computational complexity of the PLS method is dominated by the matrix factorization step, leading to a total complexity of \( O(NK^2 + K^3) \). 
\subsection{WIVE}
The computational complexity of the WIVE method can be broken down into two main steps. In the first step, the method implements the PLS estimator, which involves solving the normal equations using the measurement matrix \( A \). This step has a computational complexity of \( O(NK^2 + K^3) \), as discussed previously. In the second step, the measurement matrix is recomputed using the estimated variable vector \( \hat{\mathbf{u}} \), resulting in a new matrix $\mathbf{G}$. The variable vector \( \hat{\mathbf{u}} \) is then reestimated using the formula:
\[
\hat{\mathbf{u}} = (\mathbf{G}^T \mathbf{W} \mathbf{A})^{-1} \mathbf{G}^T \mathbf{W} \mathbf{b}~,
\]
where $\mathbf{W}$ is the weighting matrix of size \( N \times N \). The operations involved in this step include matrix multiplications and inversions. Specifically, computing \( \mathbf{G}^T \mathbf{W} \mathbf{A} \) takes \( O(NK^2) \) operations, inverting \( \mathbf{G}^T \mathbf{W} \mathbf{A} \) takes \( O(K^3) \) operations and finally, multiplying the result by \( \mathbf{G}^T \mathbf{W} \mathbf{b} \) requires \( O(NK^2) \) operations. Thus, the overall computational complexity of WIVE is dominated by these matrix operations, resulting in a total complexity of $O(NK^2 + K^3)$. This complexity is similar to that of the PLS method, with the added computation of the weighting matrix $\mathbf{W}$ and the reestimation step.
\subsection{UNet}
In a standard UNet architecture, the computational complexity of each convolutional layer is given by:
\begin{equation}
\mathcal{O}(H W C_{\text{in}} C_{\text{out}} F^2)~,
\end{equation}
where \( H, W \) are the height and width of the input image, \( C_{\text{in}} \) and \( C_{\text{out}} \) are the input and output channels of a layer, and \( F^2 \) is the filter size (assuming square kernels). Since UNet consists of an encoder-decoder structure with \( D \) layers (depth), the total computational complexity is approximately
\begin{equation}
\mathcal{O}(H W C F^2 D)~,
\end{equation}
where \( C \) is the initial number of channels, and \( D \) represents the depth of the network.

\subsection{UNet++}
UNet++ introduces additional dense skip connections and redesigned decoder pathways. Instead of direct skip connections, each feature map at level \( d \) undergoes multiple intermediate convolutional layers before reaching the decoder. This results in additional convolutional operations at each skip connection level and multiple convolutional layers within each skip pathway, effectively increasing the total number of operations per resolution level.

If each level now contains \( D \) additional intermediate convolutional layers, the total number of operations scales approximately as:
\begin{equation}
\mathcal{O}(H W C F^2 D^2)~.
\end{equation}
The additional \( D \) factor accounts for the nested skip connections, making UNet++ computationally more expensive than standard UNet. Table \ref{tab:inference_complexity} summarizes the computational complexity of each method.


\begin{table}[h]
    \centering
     \renewcommand{\arraystretch}{1.2}
    \setlength{\tabcolsep}{5pt}
    \footnotesize
    \begin{tabular}{|p{3.2cm}|p{4.8cm}|}
        \hline
        \textbf{Model} & \textbf{Inference Complexity} \\
        \hline
        PSO ML & $\mathcal{O}(TNMK)$ \\
        \hline
        PLS / WIVE & $\mathcal{O}(N K^2 + K^3)$ \\
        \hline
        UNet & $\mathcal{O}(H W C F^2 D)$ \\
        \hline
        UNet++ & $\mathcal{O}(H W C F^2 D^2)$ \\
        \hline
    \end{tabular}
    \caption{Computational complexity comparison.}
    \label{tab:inference_complexity}
\end{table}

In conclusion, PLS and WIVE have relatively low computational demands, as they primarily rely on matrix operations such as multiplications and inversions, which scale with the number of measurements and variables. In contrast, PSO-ML exhibits high complexity due to the need to compute \eqref{mlTheta} for each particle in every iteration. Deep learning models, including UNet and UNet++, involve complex convolutional networks, with computational costs increasing significantly based on image size, network depth, and the number of channels. Compared to UNet, UNet++ requires a greater number of convolutional operations, with its complexity influenced by both image dimensions and network architecture. While the inference complexity of these models may appear high, their operations can be efficiently parallelized and accelerated using Compute Unified Device Architecture (CUDA) features on GPUs. We present the experimental computation times of these methods in the following sections to validate these observations. 

\section{Experimental Results}
We divide this section into two parts reflecting the effect of the number of sources. In the first part, there is a single source, and we compare PLS, WIVE, ML, UNet, and UNet++. For the ML solver, we use PSO with the default configurations specified in the original paper \cite{kennedy1995particle}. In the second part, we focus on multiple sources and investigate the performance of deep models only.

The training data for the deep models is designed to encapsulate different scenarios, in which the noise level and source number vary. In the data, \( \sigma \) is varied \( \in \{0.5, 1, 1.5, 2, 2.5\} \) degrees, and the source number changes \( \in \{1, 2, 3, 4, 5\} \). For each combination, we generated 1000 images, making a total of 25,000 training samples. The platform carrying the EW receiver is in motion with a speed of $250$ m/s. In each scenario, its heading direction is randomly picked from the uniform distribution $\mathcal{U}(0,360)$ degrees. The measurement period and total measurement duration are randomly chosen from $\mathcal{U}(3,15)$ seconds, $\mathcal{U}(180,300)$ seconds, respectively.

Since the number of foreground pixels is much lower than the number of background pixels, the source localization problem from the perspective of semantic segmentation becomes highly sparse. To overcome this imbalanced data problem, we use a weighted Dice loss function. The weighted Dice loss is an effective modification of the standard Dice loss for handling imbalanced binary segmentation problems. By assigning appropriate weights to different classes or regions, it ensures better optimization and improves performance, particularly in applications where the positive class is significantly smaller than the negative class \cite{yeung2022unified,sugino2021loss}. It is calculated as follows:
\begin{align}
\mathcal{L}_{\text{Dice}} = 1 - \frac{2 \sum w_i y_i \hat{y}_i}{\sum w_i y_i + \sum\hat{y}_i}~.
\end{align}
Here, $y$ and $\hat{y}$ represent true source locations and estimated source locations as pixel positions, respectively. Moreover, $w_i$ represent the weight for true source locations in the images, while the background has a weight of $1$. We tried different $w_i$ values for the training and eventually preferred $w_i=1000$, which can provide a smooth learning process both models. Note that choosing a small $w_i$ might dramatically decrease the derivatives and stops learning quickly, while high $w_i$ values may cause divergence in the learning procedure.

We focus on two metrics. The first is the root mean squared-error (RMSE) value of the source localization, which is calculated using $M$ Monte Carlo runs as the following
\begin{align}
    \sqrt{\frac{1}{M}\sum_{m=1}^{M}||\mathbf{u}_m-\hat{\mathbf{u}}_m||^2}~.
\end{align}
Here, ${\mathbf{u}}_m$ and $\hat{\mathbf{u}}_m$ are the true and estimated source locations, respectively. The second metric is defined only for the multi-source localization problem, and it is the RMSE value of the source number estimation. This metric indicates the tendency of the methods to generate false alarms or missed detections. It is calculated as follows
\begin{align}
    \sqrt{\frac{1}{M}\sum_{m=1}^{M}(S_m-\hat{S}_m)^2}~.
\end{align}
Here, $S_m$ and $\hat{S}_m$ represent the true and estimated number of sources, respectively. We evaluate the performance using $M=100$ test samples for each combination of $\sigma$ and $S$, resulting in a total of $2500$ test samples. 
\subsection{Single Source Localization}
\begin{figure}[t!]
	\centering
	\includegraphics[width=1\linewidth]{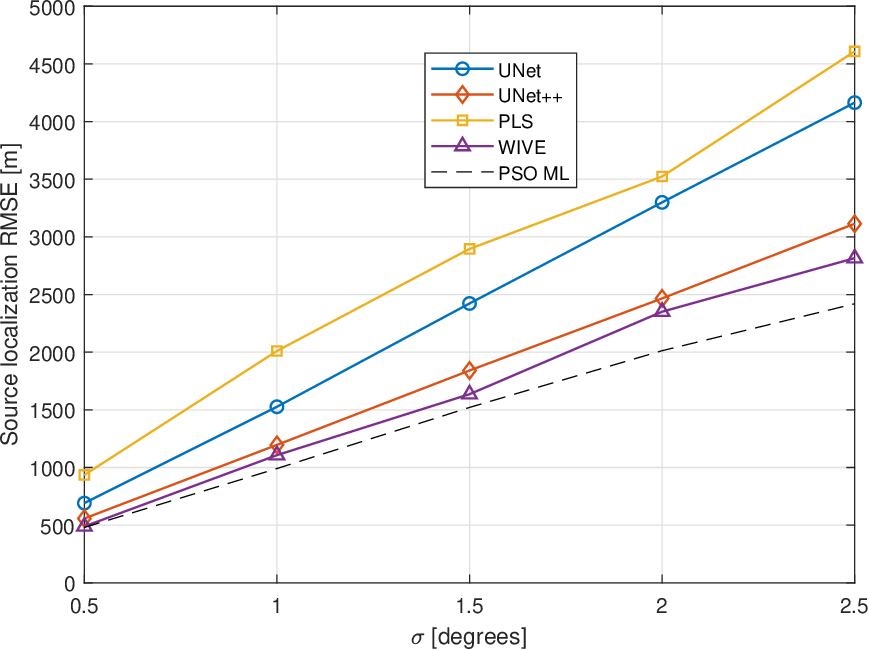}
	\caption{Source localization RMSE against $\sigma$ for single source localization scenario.}
	\label{fig:tekHedef}
\end{figure}

In Fig. \ref{fig:tekHedef}, we focus on the source localization RMSE performance of PLS, WIVE, ML, UNet, and UNet++ when there is only one source. The ML method performs the best for all noise levels because it does not rely on a pseudo-linear transform; instead, it directly targets the non-convex ML cost function. On one hand, WIVE includes an additional LS step to reduce the bias of PLS and also utilizes a weighting mechanism to emphasize measurements from shorter ranges. As a result, WIVE achieves the second-lowest RMSE. On the other hand, PLS, being the simplest method among them, depends on a pseudo-linear transform and thus has the highest RMSE. UNet outperforms PLS and provides a moderate result, while UNet++ performs similarly to WIVE but with slightly higher RMSE values. Note that the single-source localization problem can be solved mathematically, and WIVE benefits from incorporating the signal model. However, the proposed deep learning models are trained on raw input images consisting of intersecting DF lines, meaning that they do not directly utilize the signal model. This explains why outperforming a mathematical approach is a challenging task when the signal model is known, as in this case. Overall, both deep learning models outperform PLS, while U-Net++ performs similarly to WIVE. Their main advantages emerge in the multi-source localization problem, where PLS, WIVE, and ML are incapable of providing location estimates.
\begin{table}[t!]
	\begin{center}
		\caption{Average processing time of the methods when the measurement period is three seconds and measurement duration is 300 seconds.}
		\renewcommand{\arraystretch}{1.2}
		\label{tab:complexityTable3}
		\begin{tabular}{|p{1.8cm} |p{3.0cm}|p{2.5cm}|}
			\hline
			\textbf{Method} & \textit{Duration} & \textit{Platform} \\[0.5ex]
			\hline \hline
			PSO ML & $3102.270$ ms & CPU\\[0.5ex]
			\hline 
			PLS & $0.116$ ms & CPU\\[0.5ex]
			\hline
			WIVE & $0.253$ ms & CPU\\[0.5ex]
			\hline
			Preprocessing & $18.951$ ms & CPU\\[0.5ex]
			\hline
                UNet & $0.003$ ms & CUDA/GPU\\[0.5ex]
                \hline
			UNet++ & $0.003$ ms & CUDA/GPU\\
			\hline
		\end{tabular}
	\end{center}
\end{table}

Table \ref{tab:complexityTable3} presents the average processing time of the evaluated methods for a measurement period of three seconds and a total measurement duration of 300 seconds. For this purpose, a total of $100$ test samples is used and the experiments are conducted on a system equipped with an Intel Xeon w5-2445 3.10 GHz CPU and an NVIDIA RTX 4000 Ada GPU. Among the tested methods, PSO ML exhibited the longest convergence time due to its iterative nature, as it optimizes a non-convex ML function. In contrast, PLS and WIVE had significantly shorter processing times, benefiting from their closed-form analytical solutions. The proposed approach consists of two main computational stages: preprocessing and model inference. The preprocessing stage, executed on the CPU, is the most computationally intensive, as it involves extensive data preparation and transformation. In contrast, the model inference stage is highly efficient, leveraging CUDA-accelerated parallel processing on the GPU. Notably, inference runs over 100 times faster than the analytical PLS and WIVE solutions. As a result, the primary computational bottleneck of the proposed method lies in the preprocessing step, which dominates the overall execution time.

\subsection{Multi Source Localization}
In this section, we present both quantitative and qualitative results for the multi-source localization problem. Since PLS, WIVE, and ML are only applicable to single-source localization, we compare UNet and UNet++ against each other in different scenarios.
\begin{figure}[t!]
	\centering
	\includegraphics[width=1\linewidth]{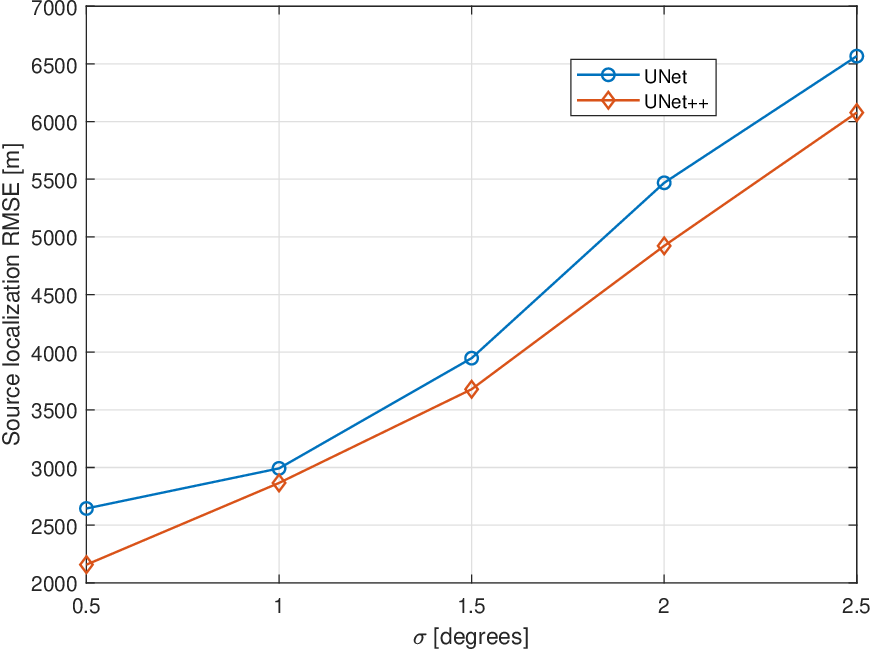}
	\caption{Source localization RMSE against $\sigma$ when the source number $S=5$.}
	\label{fig:sigmaVsLocImage}
\end{figure}

\subsubsection{Quantitative Analysis}

\begin{figure}[t!]
	\centering
	\includegraphics[width=1\linewidth]{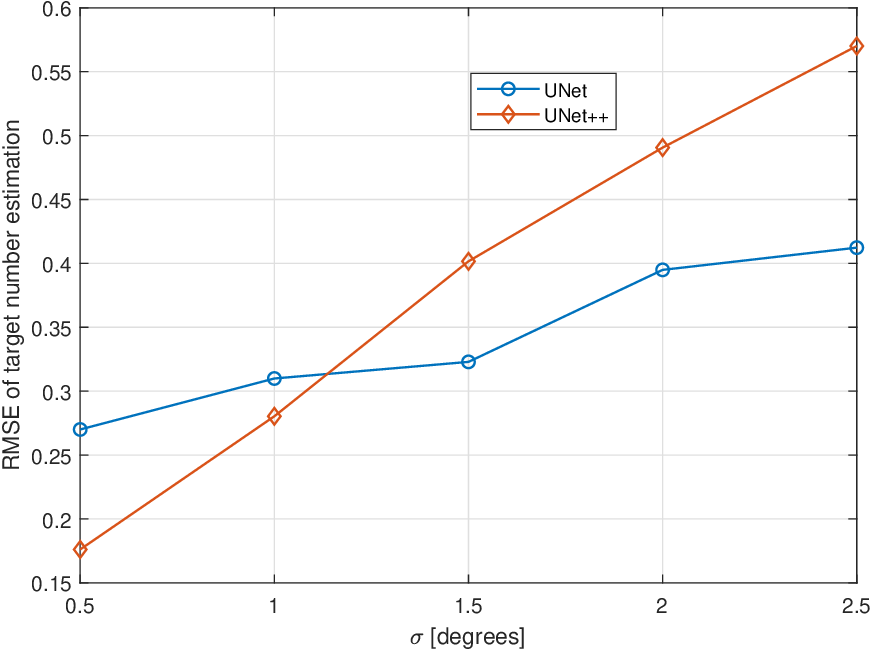}
	\caption{Source number estimation RMSE against $\sigma$ when the source number $S=5$.}
	\label{fig:sigmaVsTargetImage}
\end{figure}

In Fig. \ref{fig:sigmaVsLocImage}, we present the source localization RMSE as a function of $\sigma$ when the number of sources is $S=5$. We specifically focus on $S=5$ to evaluate the methods under the most complex scenario. We observe that UNet++ achieves better accuracy, and this performance difference persists across all noise levels. To further analyze the methods' performances in this scenario, we illustrate the source number estimation RMSE as a function of $\sigma$ in Fig. \ref{fig:sigmaVsTargetImage}, again considering $S=5$. We find that UNet++ provides more accurate source number estimations when the noise level is low. However, as the noise level increases, UNet starts to outperform UNet++. Interestingly, the performance of UNet is less affected by noise conditions, while the performance of UNet++ degrades more significantly with increasing noise levels. 

\begin{figure}[t!]
	\centering
	\includegraphics[width=1\linewidth]{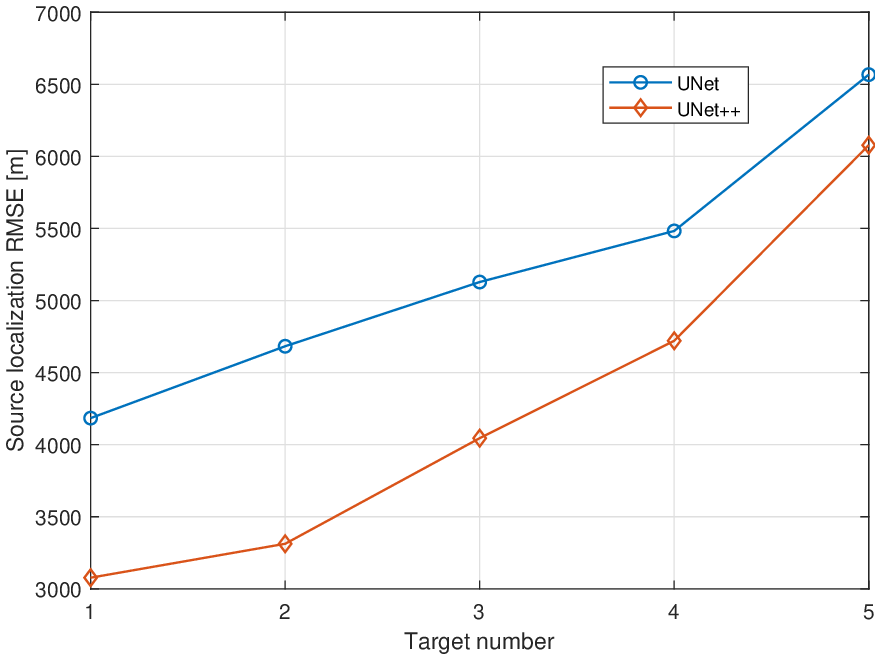}
	\caption{Source localization RMSE against the source number $S$, when $\sigma = 2.5$ degrees.}
	\label{fig:targetVsLocImage}
\end{figure}

In Fig. \ref{fig:targetVsLocImage}, we further investigate the source localization RMSE performance of the methods as a function of the number of sources, while keeping $\sigma = 2.5$ degrees. UNet++ outperforms UNet for all source numbers, and it appears that their performance is highly dependent on the number of sources. As the number of sources increases, the number of intersection points of DF lines also grows. Since the true source positions in the label image correspond to these intersection points, the models begin to confuse these intersections with actual source positions. This introduces ambiguity into the model and disrupts the learning process. 

\begin{figure}[t!]
	\centering
	\includegraphics[width=1\linewidth]{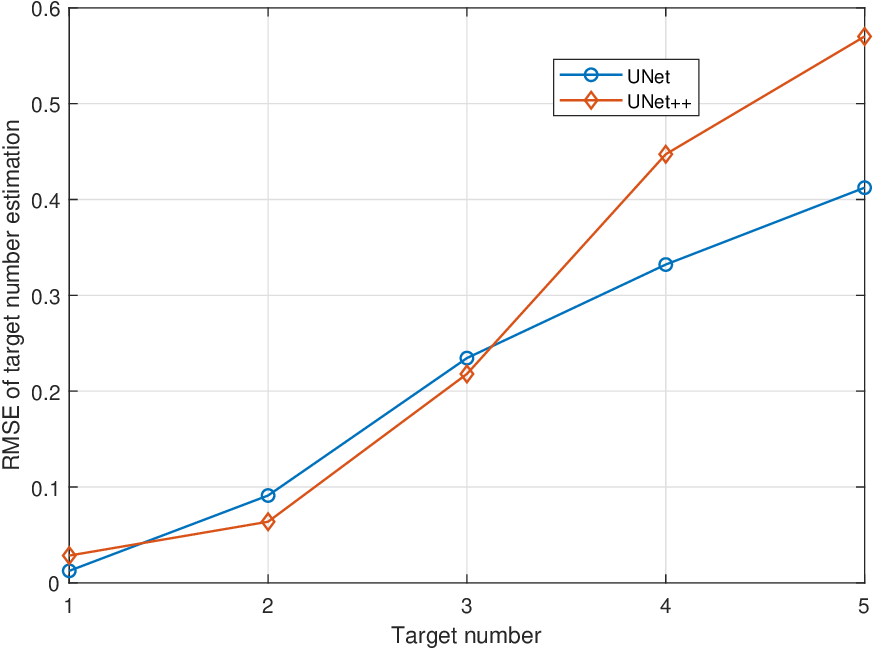}
	\caption{Source number estimation RMSE against the source number $S$, when $\sigma = 2.5$ degrees.}
	\label{fig:targetVsTargetImage}
\end{figure}

In Fig. \ref{fig:targetVsTargetImage}, we present the RMSE of the source number estimation while maintaining the noise level at $\sigma = 2.5$ degrees. The results are consistent with those in Fig. \ref{fig:sigmaVsTargetImage}, indicating that the methods perform similarly when the number of sources is low. However, as the number of sources increases, UNet++ begins to overestimate the source count, leading to higher RMSE.
\begin{figure*}[t!]
	\centering
	\includegraphics[width=1\linewidth]{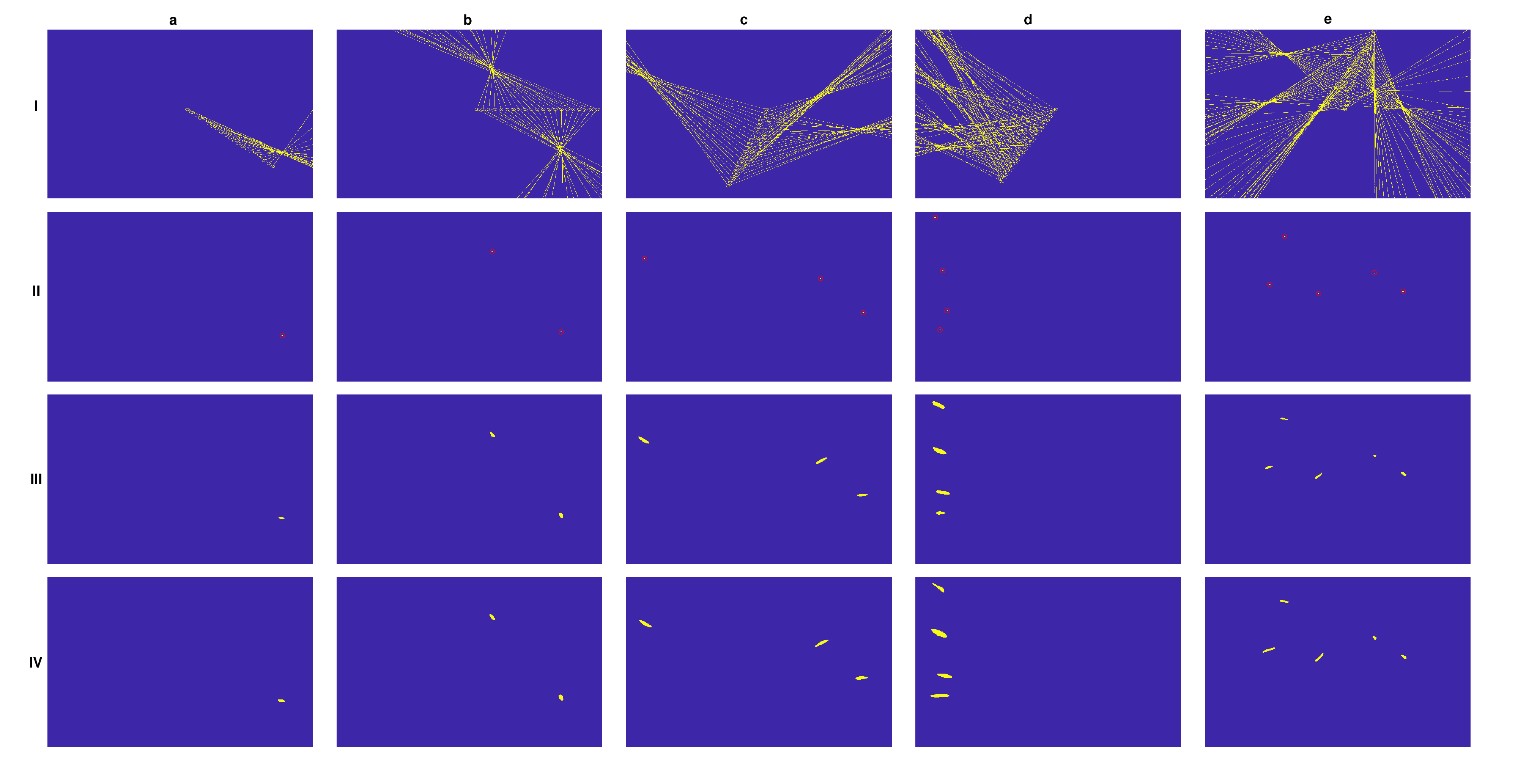}
	\caption{Five successful scenarios are presented in each column (a–e). The first and second rows, (I) and (II), represent the input and label images, respectively, while the third and fourth rows, (III) and (IV), display the outputs of UNet and UNet++, respectively.}
	\label{fig:scenarioImage}
\end{figure*}

Overall, we deduce that UNet++ provides better source localization RMSE across all noise levels and source number combinations. However, it generally estimates a higher number of sources compared to UNet. In the next section, we present example scenarios to qualitatively illustrate this observation.

\subsubsection{Qualitative Analysis}
In Fig. \ref{fig:scenarioImage}, we share five successful localization examples, organized by columns in order of increasing number of sources. The first and second rows represent the input and label images, while the third and last rows demonstrate the outputs of UNet and UNet++, respectively. We observe that both models correctly identify source locations at the center of the estimated source regions, which aligns with the label images. Upon examining the directions and shapes of these regions, we notice that they are primarily ellipses with a specific orientation. The major axes of these ellipses always point outward from the platform positions, correctly indicating the uncertainty direction of the source. Note that ellipses are commonly used in the context of source localization, as they represent possible source locations with some probability \cite{o2022practical}. Although the source locations in the label images are designed as small dots, the models detect the source locations as ellipses. Therefore, we conclude that the models learn the distribution of the source locations based on the platform geometry, which arises from the ambiguity in the intersections of DF lines.

\begin{figure}[t!]
	\centering
	\includegraphics[width=1\linewidth]{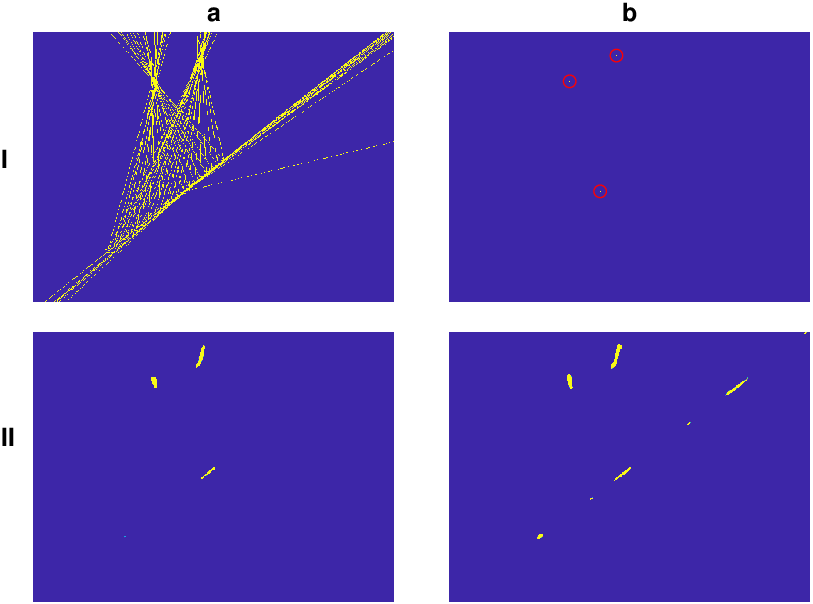}
	\caption{Failed case-I. (I-a) represents the model input, (I-b) shows the label, while (II-a) and (II-b) demonstrate the UNet and UNet++ outputs, respectively. One of the sources lies beneath the platform trajectory, and UNet++ generates false alarm.}
	\label{fig:failedCase2}
\end{figure}
\begin{figure}[t!]
	\centering
	\includegraphics[width=1\linewidth]{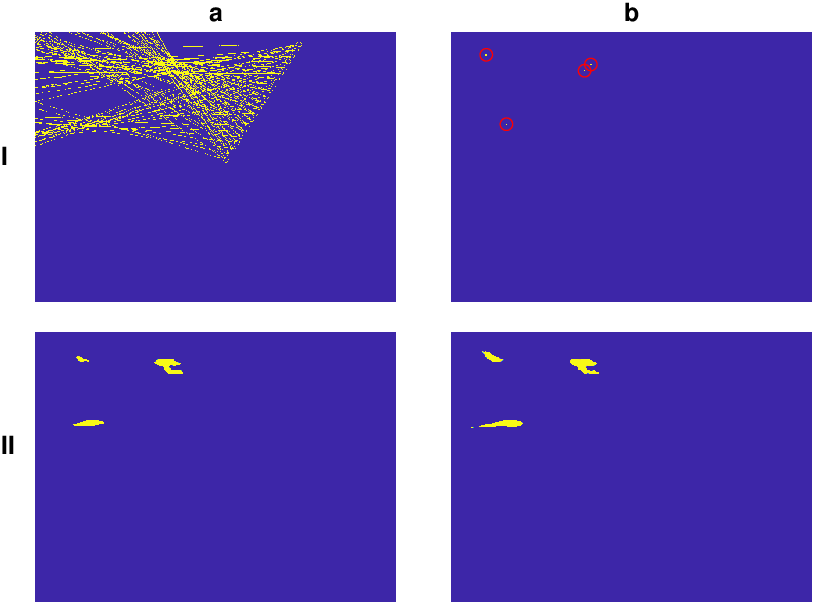}
	\caption{Failed Case-II. (I-a) represents the model input, (I-b) shows the label, while (II-a) and (II-b) demonstrate the UNet and UNet++ outputs, respectively. Two sources reside close to each other, and both models connect the two source locations into one.}
	\label{fig:failedCase3}
\end{figure}

In Fig. \ref{fig:failedCase2} and Fig. \ref{fig:failedCase3}, we present two failed cases. By analyzing these cases, potential areas for improvement can be identified, offering insights into factors that influence model performance. In the figures, (I-a) represents the model input, (I-b) shows the label, while (II-a) and (II-b) demonstrate the outputs of UNet and UNet++, respectively. In the first case, three sources exist simultaneously, and one of them lies beneath the platform trajectory, causing the platform to measure very similar AoA values over time. This scenario is undesirable for source localization methods, as it introduces potential ambiguities. In this case, UNet successfully localizes all sources. UNet++ also correctly localizes these sources, but it generates additional location estimates as well. We observe similar cases where UNet++ produces false alarms in addition to its correct location estimates. This result aligns with our findings in Fig. \ref{fig:sigmaVsTargetImage} and Fig. \ref{fig:targetVsTargetImage}, where UNet++ tends to overshoot the number of estimated sources. In the second failed case, four sources exist simultaneously, two of which are quite close to each other. We observe that both models merge these two source locations into one. Since the sources are near each other, their DF lines are intertwined, leading the models to combine their locations.


\section{Conclusions}
In this paper, we presented a novel deep learning-based approach for multi-source localization using AoA measurements. Traditional methods such as PLS, WIVE, and ML are primarily designed for single-source scenarios, whereas our approach leverages semantic segmentation models—specifically U-Net and U-Net++—to estimate multiple source locations by analyzing DF line intersections. By transforming AoA measurements into image representations and applying deep learning, we provide an efficient solution to the complex multi-source localization problem.

We evaluated our method’s performance in both single- and multi-source localization scenarios. The results show that both models perform comparably to traditional methods in single-source localization. However, in multi-source cases with varying noise levels and source densities, U-Net++ consistently achieves higher localization accuracy than U-Net. Despite its superior RMSE performance, U-Net++ tends to overestimate the number of sources, leading to occasional false alarms.

Overall, our study demonstrates the potential of deep learning models for multi-source localization in dynamic environments. Future research could focus on refining model architectures, incorporating temporal dependencies for sequential AoA data, or exploring real-world applications in electronic warfare and radar systems.

\bibliographystyle{IEEEtran}
\bibliography{main}

\end{document}